\documentclass[twocolumn,english,aps,preprintnumbers,amsmath,amssymb,superscriptaddress]{revtex4-2}
\usepackage[latin9]{inputenc}
\usepackage{amsmath}
\usepackage{graphicx}

\makeatletter
\@ifundefined{textcolor}{}
{%
 \definecolor{BLACK}{gray}{0}
 \definecolor{WHITE}{gray}{1}
 \definecolor{RED}{rgb}{1,0,0}
 \definecolor{GREEN}{rgb}{0,1,0}
 \definecolor{BLUE}{rgb}{0,0,1}
 \definecolor{CYAN}{cmyk}{1,0,0,0}
 \definecolor{MAGENTA}{cmyk}{0,1,0,0}
 \definecolor{YELLOW}{cmyk}{0,0,1,0}
}

\usepackage{dcolumn}% Align table columns on decimal point
\usepackage{bm}% bold math
\usepackage{color}

\makeatother

\usepackage{babel}

\begin{document}

\preprint{preprint(\today)}

\title{Microscopic study of the impurity effect in the kagome superconductor La(Ru$_{1-x}$Fe$_{x}$)$_{3}$Si$_{2}$}

\author{C.~Mielke III$^{\dag}$}
\email{charles-hillis.mielke-iii@psi.ch} 
\affiliation{Laboratory for Muon Spin Spectroscopy, Paul Scherrer Institute, CH-5232 Villigen PSI, Switzerland}
\affiliation{Physik-Institut, Universit\"{a}t Z\"{u}rich, Winterthurerstrasse 190, CH-8057 Z\"{u}rich, Switzerland}

\author{D.~Das}
\affiliation{Laboratory for Muon Spin Spectroscopy, Paul Scherrer Institute, CH-5232 Villigen PSI, Switzerland}

\author{J. Spring}
\affiliation{Physik-Institut, Universit\"{a}t Z\"{u}rich, Winterthurerstrasse 190, CH-8057 Z\"{u}rich, Switzerland}

\author{H.~Nakamura}
\affiliation{Institute for Solid State Physics (ISSP), University of Tokyo, Kashiwa, Chiba 277-8581, Japan}

\author{S.~Shin}
\affiliation{Laboratory for Multiscale Materials Experiments, Paul Scherrer Institut, CH-5232 Villigen PSI, Switzerland}

\author{H.~Liu}
\affiliation{Physik-Institut, Universit\"{a}t Z\"{u}rich, Winterthurerstrasse 190, CH-8057 Z\"{u}rich, Switzerland}

\author{V. Sazgari}
\affiliation{Laboratory for Muon Spin Spectroscopy, Paul Scherrer Institute, CH-5232 Villigen PSI, Switzerland}

\author{S.~J\"{o}hr}
\affiliation{Physik-Institut, Universit\"{a}t Z\"{u}rich, Winterthurerstrasse 190, CH-8057 Z\"{u}rich, Switzerland}

\author{J.~Lyu}
\affiliation{Laboratory for Multiscale Materials Experiments, Paul Scherrer Institut, CH-5232 Villigen PSI, Switzerland}

\author{J.N. Graham}
\affiliation{Laboratory for Muon Spin Spectroscopy, Paul Scherrer Institute, CH-5232 Villigen PSI, Switzerland}

\author{T.~Shiroka}
\affiliation{Laboratory for Muon Spin Spectroscopy, Paul Scherrer Institute, CH-5232 Villigen PSI, Switzerland}

\author{M.~Medarde}
\affiliation{Laboratory for Multiscale Materials Experiments, Paul Scherrer Institut, CH-5232 Villigen PSI, Switzerland}

%\author{J.-X.~Yin}
%\affiliation{Department of physics, Southern University of Science and Technology, Shenzhen, Guangdong 518055, China}

\author{M.Z. Hasan}
\affiliation{Laboratory for Topological Quantum Matter and Advanced Spectroscopy (B7), Department of Physics,
Princeton University, Princeton, New Jersey 08544, USA}
\affiliation{Princeton Institute for the Science and Technology of Materials, Princeton University, Princeton, New Jersey 08540, USA}
\affiliation{Materials Sciences Division, Lawrence Berkeley National Laboratory, Berkeley, California 94720, USA}
\affiliation{Quantum Science Center, Oak Ridge, Tennessee 37831, USA}

\author{S.~Nakatsuji}
\affiliation{Department of Physics, University of Tokyo, Bunkyo-ku, Tokyo 113-0033, Japan}
\affiliation{Institute for Solid State Physics (ISSP), University of Tokyo, Kashiwa, Chiba 277-8581, Japan}
\affiliation{Trans-scale Quantum Science Institute, University of Tokyo, Bunkyo-ku, Tokyo 113-8654, Japan}
\affiliation{Institute for Quantum Matter and Department of Physics and Astronomy, Johns Hopkins University, Baltimore, Maryland 21218, USA}
\affiliation{Canadian Institute for Advanced Research, Toronto, M5G 1Z7, ON, Canada}

\author{R.~Khasanov}
\affiliation{Laboratory for Muon Spin Spectroscopy, Paul Scherrer Institute, CH-5232 Villigen PSI, Switzerland}

\author{D.J.~Gawryluk}
\affiliation{Laboratory for Multiscale Materials Experiments, Paul Scherrer Institut, CH-5232 Villigen PSI, Switzerland}

\author{H.~Luetkens}
\affiliation{Laboratory for Muon Spin Spectroscopy, Paul Scherrer Institute, CH-5232 Villigen PSI, Switzerland}

\author{Z.~Guguchia}
\email{zurab.guguchia@psi.ch} 
\affiliation{Laboratory for Muon Spin Spectroscopy, Paul Scherrer Institute, CH-5232 Villigen PSI, Switzerland}

\begin{abstract}

We report on the effect of magnetic impurities on the microscopic superconducting (SC) properties of the kagome-lattice superconductor La(Ru$_{1-x}$Fe$_{x}$)$_{3}$Si$_{2}$ using muon spin relaxation/rotation. A strong suppression of the superconducting critical temperature $T_{\rm c}$, the SC volume fraction, and the superfluid density was observed. We further find a correlation between the superfluid density and $T_{\rm c}$ which is considered a hallmark feature of unconventional superconductivity. Most remarkably, measurements of the temperature-dependent magnetic penetration depth ${\lambda}$ reveal a change in the low-temperature behavior from exponential saturation to a linear increase, which indicates that Fe doping introduces nodes in the superconducting gap structure at concentrations as low as $x=$~0.015. Our results point to a rare example of %nodal
unconventional superconductivity in the correlated kagome lattice and accessible tunability of the superconducting gap structure, offering new insights into the microscopic mechanisms involved in superconducting order.

%The introduction of Fe neither results in magnetic order nor a spin-glass state, but rather induces a nearly negligible change in the normal state muon spin relaxation rate. 

\end{abstract}

\maketitle

\section{INTRODUCTION}

The unique kagome lattice, formed by an interwoven network of corner-sharing triangles, is a well-known playground for exploring the interplay between frustrated magnetism, electronic correlation, and topology \cite{GuguchiaCSS,JXYin2,MielkeKVS,JiangpingHu,TNeupert,MielkeTMS,BOrtizCVS,YJiang,KagomeReview,Mazin,LYe}. Thanks to the natural geometrical frustration and unique electronic structure exhibiting flat bands, van Hove singularities, and Dirac nodes, kagome lattice materials can host many fascinating physical phenomena. One of the rarest phenomena experimentally observed in kagome lattice materials is superconductivity, which has a long history of theoretically predicted exotic superconducting pairings \cite{Kiesel2013, Nandkishore,Wang2013,Qimiao}, and has been found to display competing magnetic \cite{MielkeKVS, MielkeCeRu2, KhasanovCVS, GuguchiaRVS, GuptaCVS} or otherwise unconventional \cite{MielkeLRS, GuptaCVSnpj, MielkeCDW} features. In our recent work \cite{MielkeLRS} on LaRu$_{3}$Si$_{2}$, exhibiting the highest superconducting critical temperature $T_{\rm c}$ among the known bulk kagome-lattice superconductors, we found that the $T_{\rm c}$ cannot be explained solely by electron-phonon coupling. Rather, it experiences additional enhancement from typical kagome band structure features found near the Fermi energy \cite{MielkeLRS}. This manifests a superconductivity mediated primarily by electronic correlations arising from the kagome lattice, which is an unconventional superconducting order (i.e. non-BCS). The rich interplay of competing physics in kagome systems is further complicated in LaRu$_3$Si$_2$ following the recent observation of charge order (CO) at temperatures as high as 400~K \cite{MielkeCDW}, which adds an additional electronic correlation to the normal state of this prototypical kagome superconductor.

%%%%%%%%%%%%%%%%%%%%%%%%%%%%%%%%%%%%%%%%%%%%%%%%%%%%%%%%%%%%%%
\begin{figure*}[t!]
\centering
\includegraphics[width=1.0\linewidth]{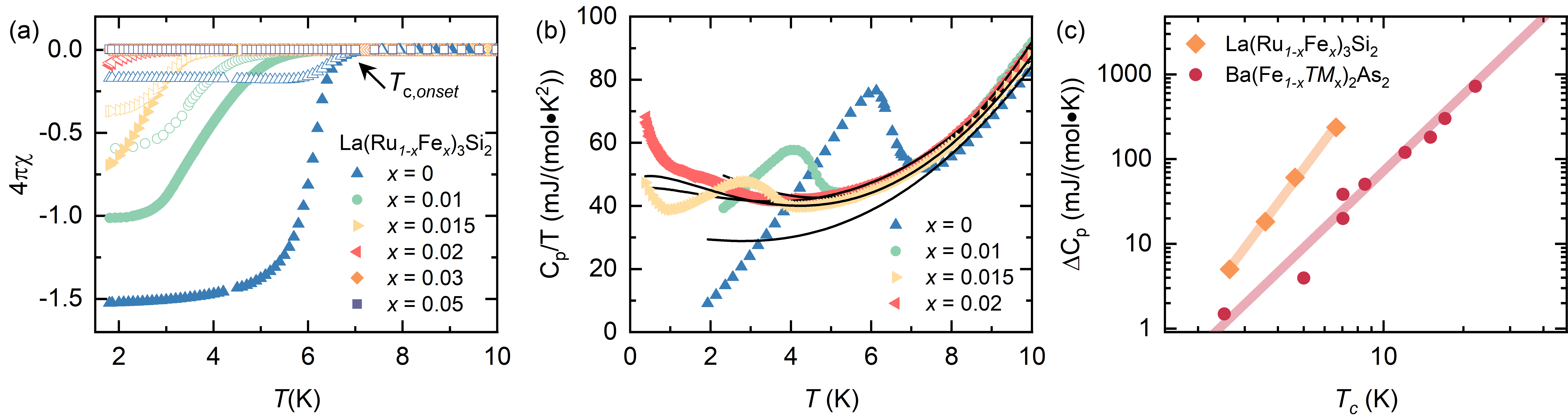}
%\vspace{0.05cm}
\caption{ %\textbf{Bulk Property Characterization.}
(a) Magnetic susceptibility vs. temperature scans for an applied field of 5 mT highlighting the suppression of superconductivity with increasing Fe concentration.  ZFC and FC conditions are denoted by closed and open symbols, respectively. The onset of the superconducting transition, defined as the intersection between the line tangent at the midpoint of the superconducting transition and the slope of the magnetization in the normal state, has been indicated by an arrow. (b) Sample specific heat divided by the temperature vs. temperature, showing clear peaks for undoped, $x$~=~0.015, and $x$~=~0.01 samples and a noticeable anomaly for the $x$~=~0.02 sample, which is however embedded in a large increase in C$_p/T$ with decreasing temperature. Fits have been overlaid as solid black lines. (c) Extracted $\Delta$C$_p$ jump vs. $T_{\rm c}$ from heat capacity measurements for each dopant concentration. Data from the current study are represented by the orange diamonds. Red triangles represent the characteristic scaling relation found for Fe-based superconductors \cite{Guguchia2016}.  It has been fitted with a linear relation on a logarithmic plot, giving a $T^{4.2}$ relation, notably steeper than the BNC $T^3$ relation. Data have been reproduced from \cite{Guguchia2016}.}
\label{fig1}
\end{figure*}
%%%%%%%%%%%%%%%%%%%%%%%%%%%%%%%%%%%%%%%%%%%%%%%%%%%%%%%%%%%%% 

Following the discovery of superconductivity in LaRu$_3$Si$_2$ \cite{discoveryLRS}, doping and impurity effect studies were performed, first on the La-site \cite{Godart1987,Escorne1994}, and more recently on the Ru-site \cite{SLi2012,BLi2016}. Previous reports on the effect of substitution of the kagome lattice-forming Ru atoms by various magnetic and non-magnetic atoms in La(Ru$_{1-x}A_{x}$)$_{3}$Si$_{2}$ revealed that $T_{\rm c}$ is rather insensitive to the element used as a dopant ($A$ = Co, Ni, Cr, Fe, Ir, Rh), except in the case of Fe \cite{SLi2012, BLi2016, SChakrabortty2023}. Fe doping causes a dramatic suppression of $T_{\rm c}$ with a critical concentration of $x_{cr, \rm Fe}\simeq$~0.036. Until now, the only known property of the superconducting state in La(Ru$_{1-x}$Fe$_{x}$)$_{3}$Si$_{2}$ is the doping-dependent evolution of $T_{\rm c}$. In light of the recently discovered tunability of the superconducting gap structure in the $A$V$_3$Sb$_5$ ($A=$~K, Rb) family \cite{GuguchiaRVS} which marks the kagome lattice as an exceptional platform to investigate competing phases and unconventional superconductivity, a microscopic understanding of the suppression of superconductivity in La(Ru$_{1-x}$Fe$_{x}$)$_{3}$Si$_{2}$ from both experimental and theoretical perspectives is required.

To shed light on the nature of the suppression of superconductivity in this prototypical kagome superconductor, we investigated the superconducting properties of La(Ru$_{1-x}$Fe$_{x}$)$_{3}$Si$_{2}$ (from $x$~=~0 to $x$~=~0.05) using a combination of complementary experimental methods such as heat capacity, magnetization, and muon spin relaxation/rotation ($\mu$SR). We observed that the introduction of Fe neither results in magnetic order nor a spin-glass state. At the same time, we provide microscopic evidence for highly tunable unconventional superconductivity in La(Ru$_{1-x}$Fe$_{x}$)$_{3}$Si$_{2}$ as evidenced by the observation of nodal superconductivity, promoted by Fe-doping, and an unconventional dependence of the superfluid density on the superconducting critical temperature.

\section{EXPERIMENTAL DETAILS AND RESULTS}

Figure 1(a) shows the temperature dependence of magnetic susceptibility $\chi$ for polycrystalline samples of La(Ru$_{1-x}$Fe$_{x}$)$_{3}$Si$_{2}$ with $x$~=~0, 0.01, 0.015, 0.02, 0.03, and 0.05, measured in an applied field of 5~mT in both zero field cooled (ZFC) and field cooled (FC) conditions, denoted by closed and open symbols, respectively. A clear suppression of both the onset of $T_{\rm c}$ and diamagnetic screening is observed as a result of Fe-doping. The sample with $x$~=~0.05 shows no diamagnetic screening, while a diamagnetic screening was clearly observed in undoped and Fe-doped samples with $x$~=~0.01, 0.015, 0.02, and a very weak diamagnetic screening in the $x$~=~0.03 sample starting just above 1.8~K, the lowest temperature achievable for our magnetization measurements.

%%%%%%%%%%%%%%%%%%%%%%%%%%%%%%%%%%%%%%%%%%%%%%%%%%%%%%%%%%%%%%
\begin{figure*}[t!]
\centering
\includegraphics[width=1.0\linewidth]{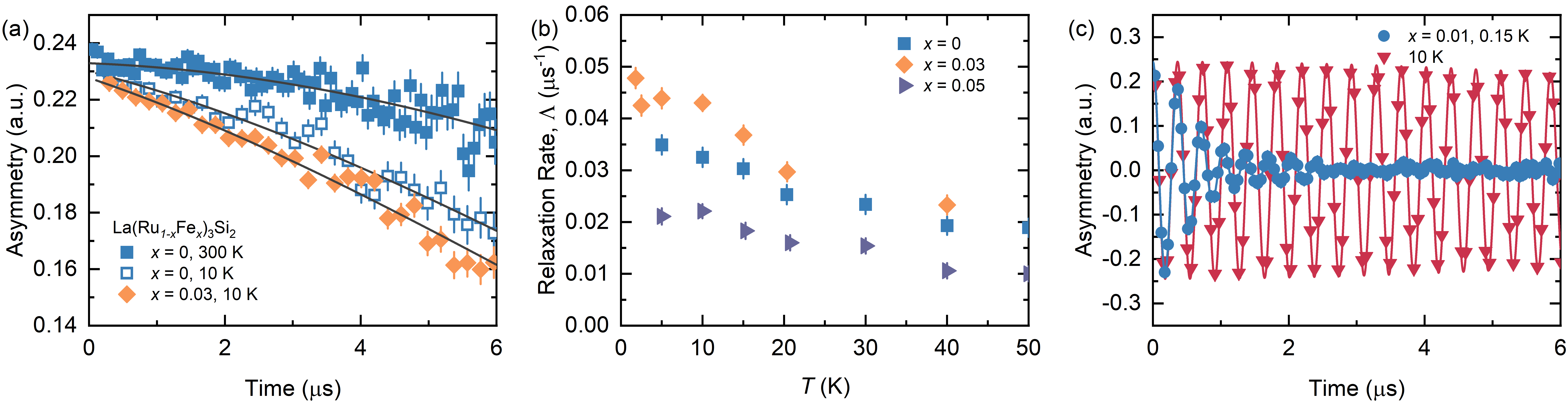}
%\vspace{0cm}
\caption{ %\textbf{Zero Field $\mu$SR Measurements.}
(a) ZF-$\mu$SR spectra, recorded for undoped and $x=$~0.03 Fe-doped samples, at 300~K and 10~K. The black line indicates the fit to the data using Eq. (2) \cite{Toyabe, Suter}. (b) Temperature dependence of the exponential relaxation rate in the undoped, $x=$~0.03, and $x=$~0.05 Fe-doped samples. The Gaussian Kubo-Toyabe was fitted with a global $\Delta_{GKT}$ parameter for each dopant concentration, which all refined to similar values (0.04-0.05~$\mu$s$^{-1}$). (c) Characteristic $\mu$SR time spectra for the $x=$~0.01 Fe-doped sample under 20~mT applied field in the normal state (at 10~K, red points) and in the superconducting state (at 0.15~K, blue points). Fits are shown as solid lines. The Fourier transformations showing field distribution can be found in the Supplementary Material \cite{SuppMat}. }
\label{fig2}
\end{figure*}
%%%%%%%%%%%%%%%%%%%%%%%%%%%%%%%%%%%%%%%%%%%%%%%%%%%%%%%%%%%%% 

In addition to diamagnetism, it is essential to check the bulk nature of superconductivity, which may be accomplished through heat capacity measurements. The heat capacity of the undoped and Fe-doped samples with $x$~=~0.01, 0.015, and 0.02 was measured and all show a peak at $T_{\rm c}$ (see Fig. 1(b)); both the value of $T_{\rm c}$ and the height of the peak are suppressed as a result of Fe-doping, indicating a reduction in superconducting volume fraction. The specific heat peak for the sample $x$~=~0.02 is severely reduced. The normal state contribution was best fitted with a model accounting for electronic correlation effects present in the sample, as previously reported \cite{SLi2011}. This equation is of the form
\begin{equation}
C = \gamma T + \beta T^3 - A T^n {\rm ln}T
\end{equation}
where  $\gamma$ is the Sommerfeld parameter, $\beta$ represents the weight of the phonon contribution, and the last term gives the correction to the Fermi liquid description, where $n$ indicates the strength of the correlations. In the case of a Fermi liquid with strong correlations $n$~=~1, while for the case of weak correlation $n$~=~3 \cite{SLi2011}. By analyzing the normal state data in the narrow temperature range just above $T_{\rm c}$ (ie. 3-10~K for $x$ = 0.02, 5-10~K for $x$~=~0.015, 5.8-10~K for $x$ = 0.01, and 7-10~K for the undoped sample), we find that $n$~$\simeq$~2. This implies that the system is moderately correlated. After subtraction of the normal state contribution, we were able to extract the jump in specific heat at $T_{\rm c}$, which is plotted against $T_{\rm c}$ in Fig. 1(c). Here, we find a linear relation when plotted on a logarithmic scale, which indicates a power-law scaling relation $\Delta$C$_p\propto T_c^{4.2}$, stronger than the $T_c^3$ scaling found for the Fe-based superconductors shown by the so-called BNC (Bud'ko-Ni-Canfield) scaling behavior \cite{Guguchia2016, Budko2009}. Additionally, the reduced peak size indicates a suppression of the superconducting volume fraction; however, the peak size in the specific heat may be affected by changes in the coupling strength or the superconducting volume fraction, while in $\mu$SR they may be extracted separately. Thus, as volume fraction is more directly probed via $\mu$SR, it will be elaborated later (see Fig.~4a). It is also important to note that the broadening of the peak may also be related to some inhomogeneity of iron doping. While there may be some distribution in iron doping, it should be distributed around some central value correlated to the nominal doping concentration. We also observe a smooth and linear evolution of superconducting properties with nominal doping concentration, indicative of the efficacy of doping. The fact that electronic correlation effects must be accounted for to fully reproduce the electronic specific heat is in line with the electronic-correlation-mediated superconductivity found in the parent material by band structure calculations \cite{MielkeLRS}. By combining the results of the magnetization and specific heat experiments, we conclude that both $T_{\rm c}$ and the superconducting (SC) volume fraction are suppressed with increasing Fe concentration.

Next, in order to identify any potential magnetic phase or ordering which may be introduced by Fe doping, zero-field $\mu$SR experiments were performed on undoped and Fe-doped samples with $x$~=~0.03 and 0.05 down to 5~K. Magnetism, if present in the samples, may enhance the muon depolarization rate in the superconducting state and falsify the interpretation of $\mu$SR results in the superconducting state. Figure 2(a) displays the zero-field ${\mu}$SR spectra for the samples $x$~=~0 and 0.03. The ZF-${\mu}$SR spectra are characterized by a weak depolarization of the muon spin ensemble and show no evidence of either long-range ordered magnetism or a spin-glass state in La(Ru$_{1-x}$Fe$_{x}$)$_{3}$Si$_{2}$. The muon spin relaxation can be fully decoupled by the application of a weak longitudinal field (LF-$\mu$SR) of 5~mT (see Supplemental Material \cite{SuppMat}), indicating that moments are static on the microsecond timescale. The ZF-$\mu$SR asymmetry spectra were fitted with a Gaussian Kubo-Toyabe function multiplied by an exponential relaxation rate as follows:
\begin{equation}
A(t) = A_0 \left[ \frac{1}{3} + \frac{2}{3} \left( 1 - (\Delta_{GKT} t) ^2 \right) e^{-\frac{1}{2}( \Delta_{GKT} t)^2 } \right] e^{- \Lambda t}
\end{equation}
where $A(t)$ is the sample asymmetry with respect to time, $A_0$ is the initial asymmetry, $\Delta_{GKT}$ is the Gaussian Kubo-Toyabe relaxation rate, and $\Lambda$ is the exponential relaxation rate \cite{Toyabe, Suter}. The Gaussian Kubo-Toyabe function reproduces the relaxation from a dense array of nuclear moments felt by the muon ensemble, while the exponential relaxation rate can be associated with the contribution of electronic origin.
The $\Delta_{GKT}$ parameter was refined as a global parameter for each sample, and the exponential relaxation rate was allowed to vary. $\Delta_{GKT}$ converged to a similar value for all three samples ($\Delta_{GKT}\simeq0.04\mu$s$^{-1}$), indicating that the nuclear moments which contribute to the Gaussian Kubo-Toyabe relaxation rate remain similar across the doping series. The temperature dependence of the exponential relaxation rate is shown in Fig. 2(b) for the undoped and $x$~=~0.03, 0.05 samples. There is only a small increase in the normal state muon-spin relaxation rate at low temperatures, which is present in the undoped as well as the Fe-doped samples. The non-monotonic dependence of the baseline and the increase in relaxation rate with decreasing temperatures deserve further attention, especially following the discovery of a high-temperature charge-ordered state \cite{MielkeCDW}, and will be the subject of future studies. Thus, the current results show that introducing small amount of Fe in La(Ru$_{1-x}$Fe$_{x}$)$_{3}$Si$_{2}$ does not introduce any additional magnetic state not already present in the parent compound.

%%%%%%%%%%%%%%%%%%%%%%%%%%%%%%%%%%%%%%%%%%%%%%%%%%%%%%%%%%%%%%
\begin{figure}[t!]
\centering
\includegraphics[width=1.0\linewidth]{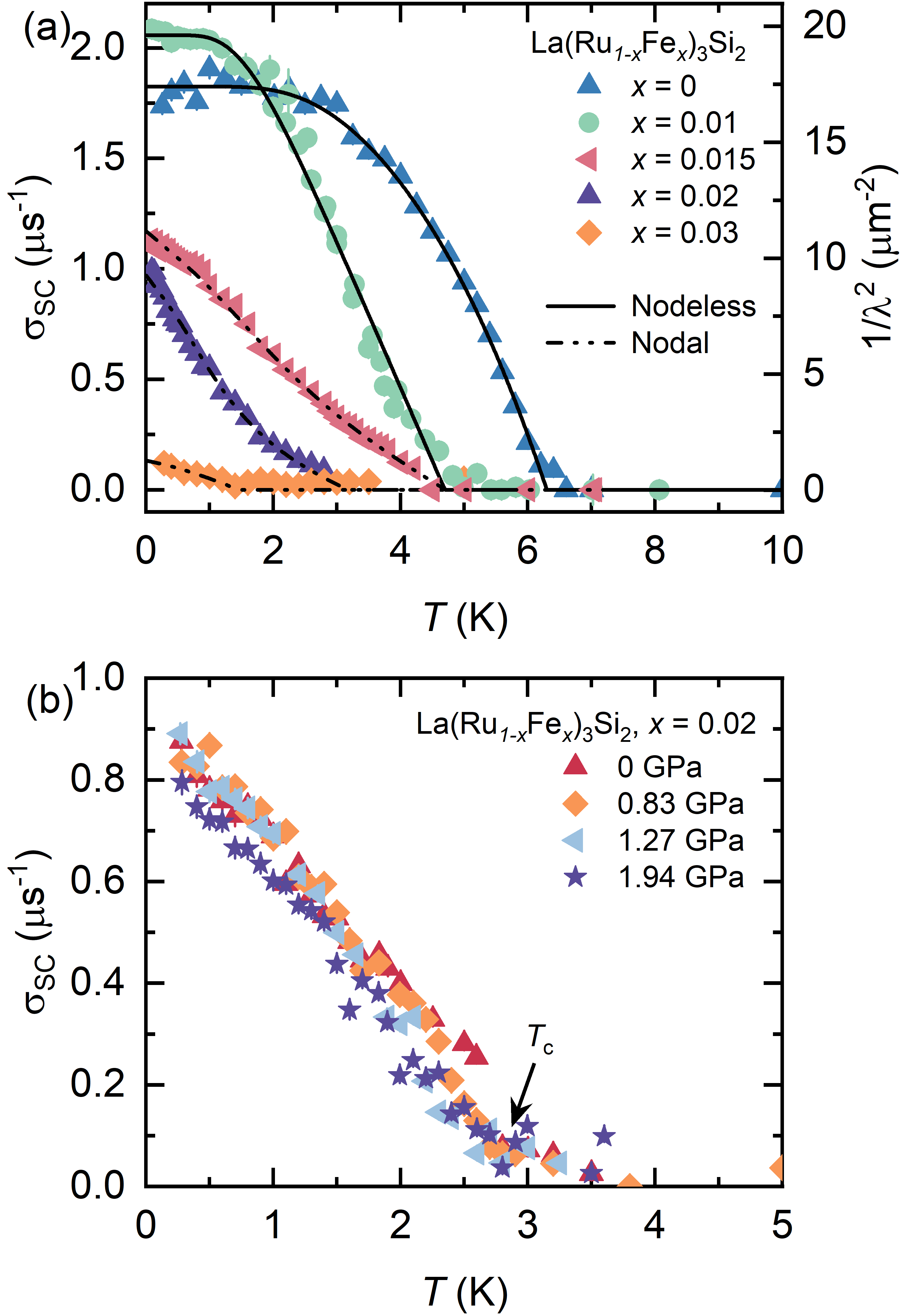}
%\vspace{0cm}
\caption{ %\textbf{$\mu$SR Measurements of Superfluid Density.}
(a) The temperature dependence of the superconducting muon-spin depolarization rate $\sigma_{\rm SC}$ (left axis) and the superfluid density (ie. $\frac{1}{\lambda^2}$) (right axis) for all doped samples including the undoped polycrystalline sample, taken from \cite{MielkeLRS}. The solid (dashed) lines correspond to fits using a model with nodeless (nodal) gap superconductivity. (b) Temperature dependence of $\sigma_{\rm SC}$  for the $x=$~0.02 Fe-doped sample under various applied hydrostatic pressures. The arrow indicates $T_{\rm c}$, which remains unchanged under pressure.}
\label{fig3}
\end{figure}
%%%%%%%%%%%%%%%%%%%%%%%%%%%%%%%%%%%%%%%%%%%%%%%%%%%%%%%%%%%%% 

Once it was established that no enhancement of muon depolarization rate was introduced by Fe doping, the suppression of the superconducting state could be explored microscopically using transverse field (TF) $\mu$SR. We probe the SC order parameter via the superfluid density $\sigma_{\rm SC}~\propto~\frac{1}{\lambda^2}~\propto~n_s$, where $\sigma_{\rm SC}$ is the superconducting muon spin depolarization rate. The absence of long-range magnetic order or a spin-glass state in La(Ru$_{1-x}$Fe$_x$)$_3$Si$_2$ implies that the increase of the TF relaxation rate below $T_{\rm c}$ is attributed entirely to the vortex lattice. As an example, the TF-$\mu$SR spectra for the $x$~=~0.01 sample, measured in an applied magnetic field of 20~mT above (10~K) and below (0.15~K) the SC transition temperature $T_{\rm c}$, are shown in Figure~2c. Above $T_{\rm c}$ the oscillations show a small relaxation rate $\sigma_{\rm nm}$ due to the random local fields from the nuclear magnetic moments. At 0.15~K, the relaxation rate increases due to the presence of a nonuniform local field distribution as a result of the formation of a flux-line lattice (FLL) in the SC state (see the Supplemental Material \cite{SuppMat}, as well as Refs. \cite{Brandt, Suter, Tinkham, carrington} therein). The base-$T$ value of the superfluid density increases for the $x$~=~0.01 sample, before decreasing rapidly for $x$~=~0.015, $x$~=~0.02, and $x$~=~0.03 samples (see Fig.~3(a)). Despite the non-monotonic behavior in superfluid density, the $T_{\rm c}$ decreases monotonically with increasing Fe-dopant concentration (doping evolution of $T_{\rm c}$ is plotted in Fig.~4(a)). We also find a constant decrease in SC volume fraction with increased doping concentration which perfectly follows the $\frac{T_{\rm c}}{T_{\rm c,~pure}}$ dependence at low concentrations before diverging at $x=$~0.015, $x=$~0.02  and $x=$~0.03, see Fig.~4(a). 

More importantly, the temperature dependence of the superfluid density suggests a change in the SC gap structure: in undoped and $x$~=~0.01 samples, the temperature dependence of the superfluid density $n_s~\propto~\frac{1}{\lambda^2}~\propto~\sigma_{\rm SC}$ saturates below $\simeq2$~K and 1.5~K, respectively, and has been well-modeled with the $s$-wave superconducting gap symmetry \cite{MielkeLRS}. In contrast, in the $x$~=~0.015, $x$~=~0.02, and 0.03 samples, the dependence of $\sigma_{\rm SC}$ increases linearly with decreasing temperature, indicating the presence of a SC gap which has nodes. This points towards a crossover from fully gapped to nodal SC gap structure with increasing Fe-dopant concentration. The introduction of Fe doping seems to have the inverse effect of hydrostatic pressure on other kagome superconductors such as KV$_3$Sb$_5$ and RbV$_3$Sb$_5$ \cite{GuguchiaRVS}, which exhibit a nodal SC gap symmetry under ambient conditions and become fully gapped under hydrostatic pressure. 

The nodal superconducting state introduced by Fe doping is remarkably stable, demonstrated by additional $\mu$SR studies under pressure. The application of hydrostatic pressures up to 2~GPa was achieved, and the temperature dependence of $\sigma_{\rm SC}$ for $x=$~0.02 showed no saturation down to the lowest temperatures (see Fig. 3(b)). In fact, the data for all four pressures probed by $\mu$SR nearly perfectly overlap, as seen in Fig. 3(b). This implies that the application of hydrostatic pressure has almost no effect on the superfluid density or $T_{\rm c}$ in the full pressure range, much like in the parent compound LaRu$_3$Si$_2$ \cite{MielkeLRS}. This indicates that nodal  superconductivity is the energetically preferred groundstate at $x$~=~0.015, $x=$~0.02 and $x$~=~0.03 concentrations. 

From the $\mu$SR measurements, we are able to construct the phase diagram with respect to doping concentration, which is shown in Figure 4a. We observe a non-monotonic behavior of the superfluid density with Fe doping, while simultaneously identifying a continuous decrease of superconducting $T_{\rm c}$ with Fe dopant concentration. Thus we find that superconductivity is suppressed steadily from three different directions: in terms of $T_{\rm c}$, volume fraction, and superfluid density. The anomalous increase of 20\% in superfluid density at $x$~=~0.01 Fe-doping concentration stands out however. In the context of other superconductors \cite{Uemura1,Uemura2,Shengelaya}, we see in Fig.~4(b) that the $T_{\rm c}$ and superfluid density (ie. $\frac{1}{\lambda^2}$) values place the undoped and Fe-doped samples with $x$~=~0.01, $x$~=~0.015 and $x$~=~0.02 all neatly along the line formed by other kagome systems \cite{GuguchiaRVS} such as CeRu$_2$ and CsV$_3$Sb$_5$ as well as moderately correlated quasi-2D transition metal dichalcogenides (TMDs) \cite{GuguchiaMoTe2,GuguchiaNbSe2}. This suggests a universal scaling relation common among the kagome superconductors, similar to that found in TMDs as well. The data point for the $x$~=~0.03 sample lies near KV$_3$Sb$_5$ and RbV$_3$Sb$_5$ and on the same line as electron-doped cuprates, but all members in this doping series lie far from the conventional superconductors and show correlation between the superfluid density and $T_{\rm c}$, which is considered a hallmark feature of unconventional superconductivity.

%%%%%%%%%%%%%%%%%%%%%%%%%%%%%%%%%%%%%%%%%%%%%%%%%%%%%%%%%%%%%%
\begin{figure*}[t!]
\centering
\includegraphics[width=1.0\linewidth]{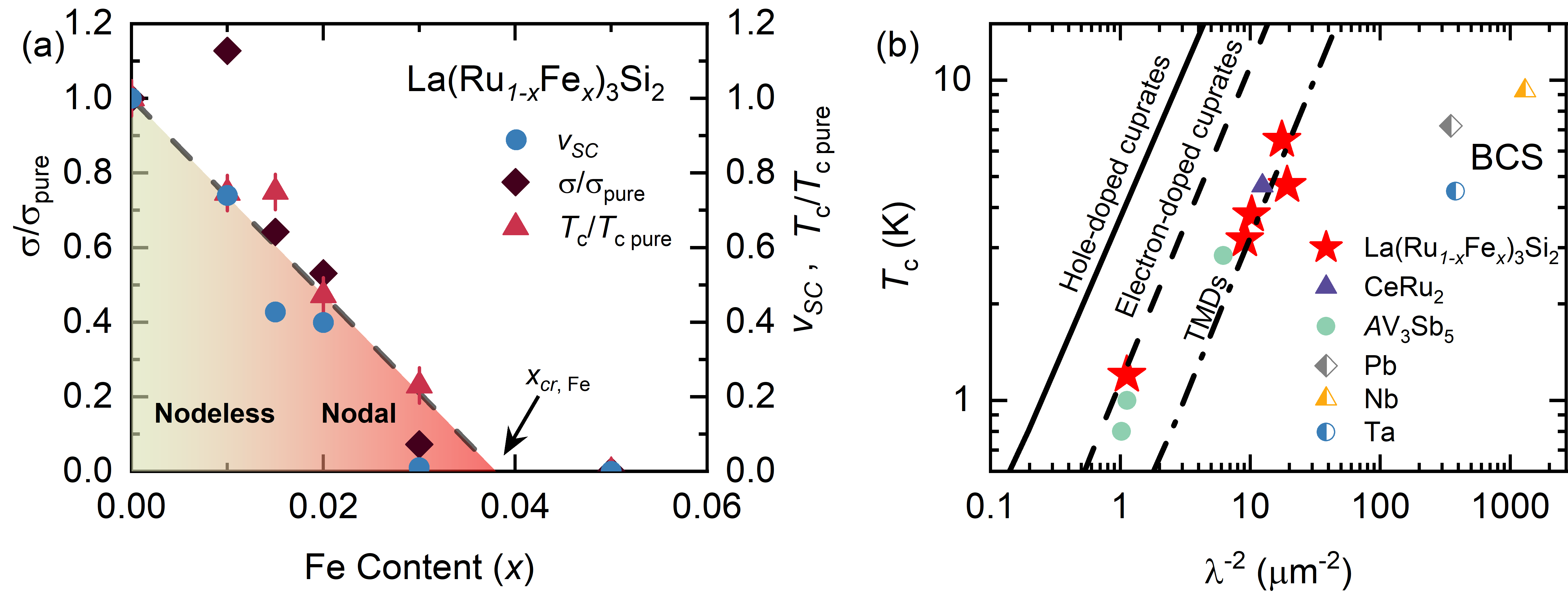}
%\vspace{-2.8cm}
\caption{ (a) The doping evolution of the superconducting muon-spin depolarization rate $\sigma_{\rm SC}$, normalized superconducting $T_{\rm c}$, and superconducting volume fraction $V_{\rm SC}$, extracted from the $\mu$SR results. The light green-to-red shaded region indicates the crossover around $x=$~0.015 from nodeless superconducting gap structure (light green) to a nodal structure (red). (b) The $T_{\rm c}$ vs. superfluid density ratio has been plotted for the different dopant concentrations of La(Ru$_{1-x}$Fe$_{x}$)$_3$Si$_2$ compared with other known conventional and unconventional superconductors \cite{Uemura1,Uemura2,Shengelaya,GuguchiaMoTe2,GuguchiaNbSe2,GuguchiaRVS}. While all dopant concentrations lie far from the elemental conventional superconductors, they fall almost entirely along the same line as several example transition metal dichalcogenides \cite{GuguchiaMoTe2,GuguchiaNbSe2} and other kagome superconductors \cite{GuguchiaRVS}. Only the 3\% Fe-doped sample falls a bit away from this line, and falls instead on the same line as the electron-doped cuprates, similarly to RbV$_3$Sb$_5$ and KV$_3$Sb$_5$ under ambient pressure.}
\label{fig4}
\end{figure*}
%%%%%%%%%%%%%%%%%%%%%%%%%%%%%%%%%%%%%%%%%%%%%%%%%%%%%%%%%%%%% 
 
\section{DISCUSSION}
 
There is a long history of predicting \cite{Kiesel2013,Nandkishore,Wang2013} different types of electronic instabilities on the kagome lattices at select fillings (e.g. 5/4 electrons per band) such as charge-density wave order, bond density wave state, chiral spin density wave states, and superconductivity with $d+id$- or $f$-wave superconductivity. The experimental realization of these exotic superconducting pairings has been largely missing until recently, following the discovery of the $A$V$_3$Sb$_5$ ($A$~=~K, Rb, Cs) family, which has unleashed an avalanche of interest in kagome systems. A nodal superconducting pairing has been found experimentally in KV$_3$Sb$_5$ and RbV$_3$Sb$_5$, and can be tuned to nodeless $s$-wave symmetry by the application of hydrostatic pressure \cite{GuguchiaRVS} and by doping \cite{YZhong}. The crossover from nodal to nodeless pairing is correlated with the establishment of the optimal superconducting region of the phase diagram, which corresponds to full suppression of charge order in KV$_{3}$Sb$_{5}$ and partial suppression of charge order in RbV$_{3}$Sb$_{5}$. The essential findings of this paper are the observation of the unconventional scaling of the superfluid density with respect to $T_{\rm c}$, and the crossover from nodeless to a superconducting gap structure with nodes, promoted by Fe-doping in La(Ru$_{1-x}$Fe$_{x}$)$_{3}$Si$_{2}$. Thus, in the present case of La(Ru$_{1-x}$Fe$_{x}$)$_{3}$Si$_{2}$, we have uncovered yet another highly tunable unconventional kagome superconductor, and are able to introduce nodes in the superconducting gap structure with the introduction of a small quantity of Fe doping without introducing a competing magnetic state. It is furthermore interesting to note that a high-temperature charge order has been uncovered in the La(Ru$_{1-x}$Fe$_{x}$)$_{3}$Si$_{2}$ system \cite{MielkeCDW}; however, the onset temperature lies around 400~K in the parent, undoped compound LaRu$_{3}$Si$_{2}$, which exhibits nodeless superconductivity and a $T_{\rm c}$~=~7~K \cite{MielkeLRS}, similar to CsV$_3$Sb$_5$ once charge order has been fully suppressed \cite{GuptaCVS}. Therefore, it seems that the Fe-doping causes the changes in the Fermi surface without the presence of a competing state. Whether the SC gap topology is well-ordered (ie. $d$-wave superconductivity, where the nodes occur at $\frac{\pi}{2}$ spacing) or the nodes are ``accidental" \cite{GuguchiaNature,Fernandes2012}  is difficult to determine with $\mu$SR. Further investigations are desirable to distinguish between the two.

\section{CONCLUSION}

In conclusion, we have studied the effect of Fe-doping on the superconducting and normal state properties of the prototypical kagome superconductor LaRu$_3$Si$_2$. The introduction of iron on the ruthenium site neither precipitates a magnetic state with long-range order nor introduces a novel magnetic state not already present in the parent compound. However, we have observed a strong suppression of superconductivity, which is manifested by the suppression of the volume fraction, suppression of $T_{\rm c}$, and suppression of the superfluid density. We further show an unconventional dependence of the superfluid density on the superconducting critical temperature. Most importantly, we find a crossover from fully-gapped superconductivity in the parent compound to a superconductivity with nodes by $x$~=~0.015 Fe-substitution, less than half the critical doping concentration required to fully suppress superconductivity. This study will serve to stimulate useful discussion regarding the nature of kagome superconductivity and the delicate balance it holds between competing orders.

%%%%%%%%%%%%%%%%%
\section{Acknowledgments}~
The ${\mu}$SR experiments were carried out at the Swiss Muon Source (S${\mu}$S) Paul Scherrer Insitute, Villigen, Switzerland using the surface muon beamlines leading to the GPS instrument and high-field HAL-9500 instrument, equipped with a BlueFors vacuum-loaded cryogen-free dilution refrigerator. Additional ${\mu}$SR experiments under hydrostatic pressure were carried out using the decay beamline and instrument GPD, also at the Swiss Muon Source (S${\mu}$S). C.M. acknowledges useful discussions with E. Pomjakushina and thanks her for her expertise and experience in synthesis. Z.G. acknowledges useful discussions with Dr. Robert Johann Scheuermann and Professor Jia-Xin Yin. Z.G. acknowledges support from the Swiss National Science Foundation (SNSF) through SNSF Starting Grant (No. TMSGI2${\_}$211750). We also acknowledge the Swiss National Science Foundation grant No. 200021${\_}$188706. The heat capacity measurements were carried out on the PPMS device of the Laboratory for Multiscale Materials Experiments, Paul Scherrer Institute, Villigen, Switzerland and on the PPMS device at the UZH in Z\"{u}rich, Switzerland (SNSF grant No. 20-175554). The magnetization measurements were carried out on the MPMS device of the Laboratory for Multiscale Materials Experiments, Paul Scherrer Institute, Villigen, Switzerland (SNSF grant no. 206021${\_}$139082) and on the MPMS3 device at the UZH in Z\"{u}rich, Switzerland (SNSF grant No. 206021-150784). Work at the University of Tokyo was supported by the JST-Mirai Program (Grant No. JPMJMI20A1), JST-CREST (Grant No. JPMJCR18T3), and JST-ASPIRE (Grant No. JPMJAP2317). M.Z.H. acknowledges support from the US Department of Energy, Office of Science, National Quantum Information Science Research Centers, Quantum Science Center at ORNL and Laboratory for Topological Quantum Matter at Princeton University.\\

\clearpage

\renewcommand{\figurename}{Supplementary Figure S}
\maketitle
\section{Supplementary Information}

\textbf{General remark}: Here, we concentrate on muon spin rotation/relaxation/resonance  ($\mu$SR) \cite{Sonier,Brandt,GuguchiaNature} measurements of the magnetic penetration depth $\lambda$ in La(Ru$_{1-x}$Fe$_{x}$)$_{3}$Si$_{2}$, which is one of the fundamental parameters of a superconductor, since it is related to the superfluid density $n_{s}$ via 1/${\lambda}^{2}$ = $\mu_{0}$$e^{2}$$n_{s}/m^{*}$ (where $m^{*}$ is the effective mass). Most importantly, the temperature dependence of ${\lambda}$ is particularly sensitive to the structure of the SC gap. Moreover, zero-field ${\mu}$SR is a very powerful tool for detecting static or dynamic magnetism in exotic superconductors, because very small internal magnetic fields are detected in measurements without applying external magnetic fields.\\ 

\textbf{Sample preparation/Synthesis}: These samples were synthesized from high-purity Si lump (purity 99.999+\%, Alfa Aesar), low-oxygen vacuum remelted Fe lump (purity 99.99\%, Alfa Aesar), Ru pellets (purity 99.95\%, Alfa Aesar), and La ingot (purity 99.9\%, Alfa Aesar) via arc melting using Zr pellets as an oxygen getter. The melted buttons were flipped three times to ensure melt homogenization. In order to suppress the second phase LaRu$_2$Si$_2$, an additional 15\% Ru was added to each melt (as mentioned in previous research \cite{SLi2011, SLi2012, BLi2016}) in order to attain nominal compositions as La(Ru$_{1-x}$Fe$_{x}$)$_{3.45}$Si$_{2}$. The arc melted samples were then shattered and a small portion was ground, upon which was performed powder X-ray diffraction (PXRD) using the D8 Advance Powder X-ray Diffractometer (BRUKER) with Cu K$_{\alpha}$ radiation. The collected PXRD patterns were refined using the FullProf suite \cite{Rodriguez}, using a multi-phase refinement to capture the presence of additional Ru and any second-phase LaRu$_2$Si$_2$ which might have formed. In all samples used for $\mu$SR, the presence of second phases (other than Ru, of which an extra 15\% was intentionally introduced) refined to a value $\leq$ 5\%.\\

\textbf{Magnetization Measurements}: Magnetization measurements were performed on all the doped concentrations, and have been converted to true $\chi$. The magnetization data was treated in the following way: 
\begin{equation}
	\chi = \frac{4 \pi M \rho}{m B}
\end{equation}
where $M=$ magnetic moment [emu], $\rho =$ sample density [g/cm$^3$], $m =$ sample mass [g], $B =$ applied magnetic field [Oe]. %(the demagnetization factor was not considered because all experiments were performed on several polycrystalline chunks of as-grown polycrystalline sample).

We measured undoped ($x$~=~0), $x$~=~0.01, $x$~=~0.015, $x$~=~0.02, $x$~=~0.03, and $x$~=~0.05 Fe-doped La(Ru$_{1-x}$Fe$_{x}$)$_{3}$Si$_{2}$ on the MPMS3 with VSM option at the UZH Irchel and on the MPMS-XL SQUID at the PSI in the LMX laboratory. On the MPMS3 at UZH Irchel, we glued a small fragment of the as-prepared sample to a quartz sample stick using GE varnish. The samples were shattered from the arc-melted button%, and the fragments used were always much longer in the $z$-axis (along the direction of $\Vec{B}_{app}$, lending itself to a demagnetization factor $\simeq$~1). 
. For the SQUID measurements on the MPMS at PSI, several irregular chunks of sample were mounted in a capsule and fixed in place with cotton fluff to prevent sample movement during scanning or field application. 

%%%%%%%%%%%%%%%%%%%%%%%%%%%%%%%%%%%%%%%%%%%%%%%%%%%%%%%%%%%%%%
\begin{figure}[t!]
	\centering
	\includegraphics[width=1.0\linewidth]{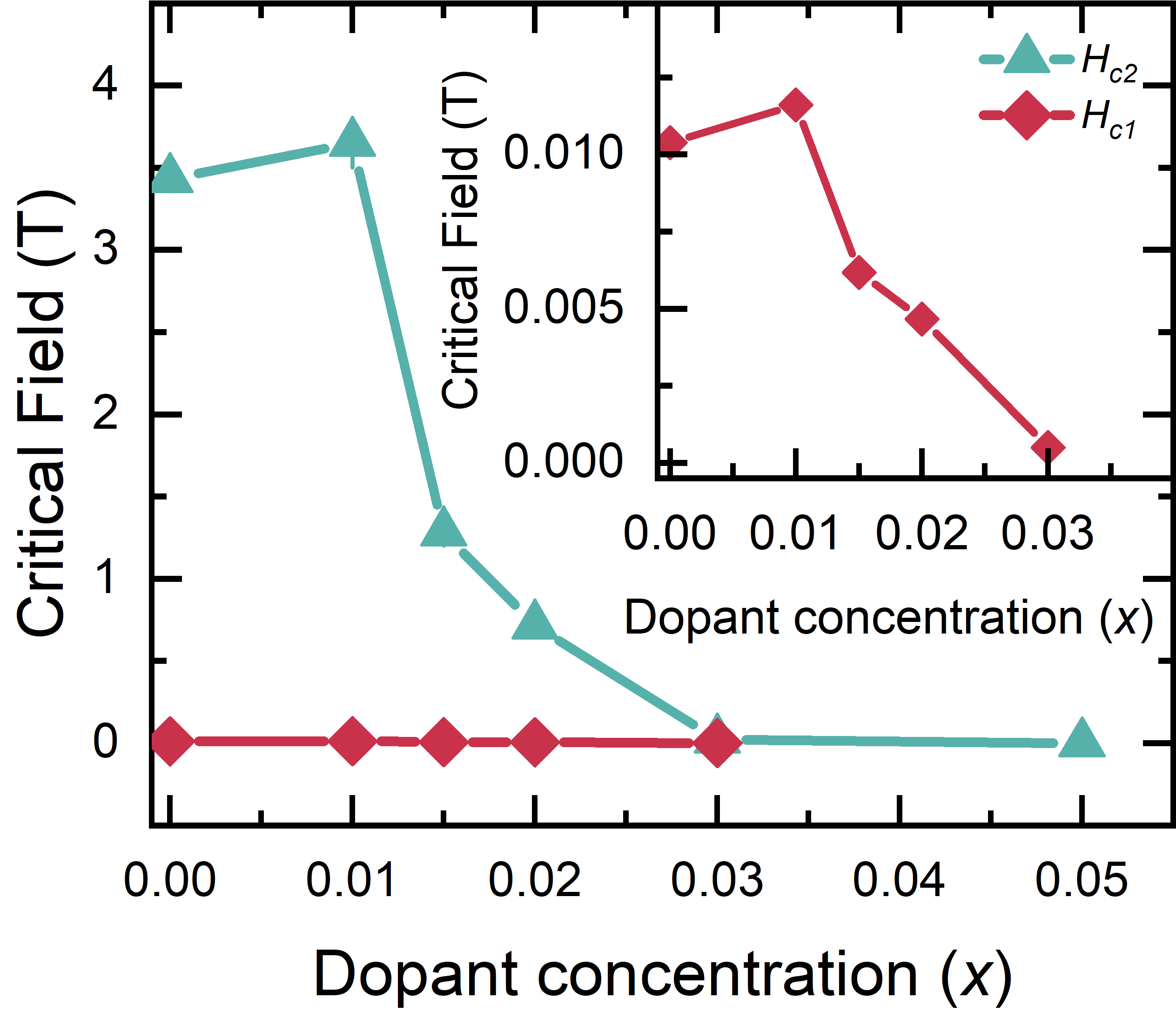}
	%\vspace{-2.8cm}
	\caption{{\bf Critical fields in the superconducting state of La(Ru$_{1-x}$Fe$_x$)$_3$Si$_2$.} The upper critical field $H_{\rm c2}$ was measured via magnetization for various Fe-dopant concentrations. The lower critical field $H_{\rm c1}$ has been calculated from magnetization and $\mu$SR measurements (see Eq.s 4\&5).}
	\label{fig1}
\end{figure}
%%%%%%%%%%%%%%%%%%%%%%%%%%%%%%%%%%%%%%%%%%%%%%%%%%%%%%%%%%%%% 

For all of the superconducting samples, $H_{\rm c2}$ was taken as the intersection between the lines formed by the normal state trend and the slope at the midpoint of the superconducting transition. 

The lower critical field $H_{\rm c1}$ may be calculated for a type-II superconductor utilizing the following formula:
\begin{equation}
	\mu_0 H_{\rm c1} = \frac{\Phi_{0}}{4\pi\lambda^{2}} \left[ ln(\kappa) + 0.5 \right]
\end{equation}
where $\Phi_{0}=$~2.07$\times$10$^{-3}$~T$\mu$m$^2$ is the flux quantum, $\kappa=\frac{\lambda}{\xi}$ where $\lambda$ is the superconducting penetration depth and $\xi$ is the correlation length, which may be found by 
\begin{equation}
	\xi = \sqrt \frac{\Phi_{0}}{2\pi H_{\rm c2}}
\end{equation}
calculated from the upper critical field $H_{\rm c2}$. A full discussion and derivation may be found in \cite{TShang2019}. It is important for a muSR experiment to be in the superconducting vortex state, ie. $H_{\rm c1}<H_{\rm app}<H_{\rm c2}$.

To extrapolate $H_{\rm c2}$ at 0~K, the critical field was measured at various temperatures and plotted together for each dopant concentration. From this, a linear extrapolation to 0~K could be made by finding the intercept. The superconducting penetration depth $\lambda$ was calculated from $\mu$SR measurements of the increased relaxation rate (see Equation 11 in the section \textbf{Transverse Field $\mu$SR}). We find that the $H_{\rm c1}$ and $H_{\rm c2}$ both increase from undoped to $x=$~0.01 Fe-doped sample, before decreasing for all additional dopant concentrations. It should be noted that the calculated $H_{\rm c1}$ was always less than the field $\mu$SR was measured in, with a maximum $H_{\rm c1}=$~0.0116~T in the $x=$~0.01 Fe-doped sample.\\

\textbf{Heat Capacity Measurements}: Heat capacity was measured on the PPMS at the UZH-Irchel and at the Paul Scherrer Institut using the He$^3$ option on the PPMS-14T in the LMX laboratory. Heat capacity was measured by the relaxation method. For both samples, first an addenda was taken on a calibrated heat capacity puck, where for the addenda measurement a small amount of Apiezon N thermal grease was smeared on the puck. The sample was then applied and stuck to the puck with Apiezon grease, to achieve good thermal contact between the sample and the puck.

The heat capacity was fitted first with the standard Debye model, by plotting sample heat capacity divided by temperature vs. temperature squared, and then extracting the slope and intercept in the temperature range immediately preceding the sample superconducting transition. This then provides the parameters in the Debye equation
\begin{equation}
	\frac{C_p}{T} = \gamma_n + \beta T^2
\end{equation}
where  $\gamma_n$ is the Sommerfeld parameter. This model, however, achieved poor agreement with the data, and resulted in a large residual lump in specific heat in the $x$~=~0.01, $x$~=~0.015, and $x$~=~0.02 Fe-doped samples. For this reason, we turned to a higher order expansion which accounts for higher frequency phonon modes, namely we fitted with the equation 
\begin{equation}
	\frac{C_p}{T} = \gamma_n + \beta T^2 + \eta T^4
\end{equation}
where the last two terms come from the phonon contributions. This model achieved slightly better agreement, but still resulted in a large residual lump in the $x$~=~0.02 Fe-doped sample in the vicinity of the superconducting transition, and diverged at higher temperatures. Finally, we turned to another fitting function which would indicate some electronic correlation effect present in the sample, as mentioned by \cite{SLi2011}. This equation is of the form
\begin{equation}
	C_p = \gamma_n T + \beta T^3 - A T^n lnT
\end{equation}
where $n$ indicates the strength of the correlations, varying from  $n$~=~1~-~3 \cite{SLi2011}. We allowed $n$ to vary as a free parameter, but it converged to $n$~$\simeq$~2 for fits performed on the normal state data immediately before the SC transition. From this, we extracted parameters in surprising agreement with previous measurements \cite{SLi2011}.
\begin{table}
	\begin{tabular}{ c c c c}
		\hline
		\multicolumn{4}{|c|}{Parameters} \\
		\hline
		$x_{\rm Fe}$ & $\gamma$& $\beta$ & A\\
		\hline
		undoped \cite{SLi2011}   & 36.8    &1.416&   3.61\\
		undoped &   29.06  & 1.52   & 4.2\\
		0.01 &54.1 & 2.47&  9.08\\
		0.015 & 46.6 & 2.02 &  6.99\\
		0.02 & 51.31 & 2.12 &  7.52\\
		\hline
	\end{tabular}
	\caption{ \textbf{Specific Heat Parameters.}
		The parameters from the refinements using Eq.(6) of the specific heat measurements are listed above for the undoped, $x=$~0.01, $x=$~0.015, and $x=$~0.02 Fe-doped samples. For comparison, the refinement parameters from \cite{SLi2011} are provided as well.}
\end{table}\\

\textbf{${\mu}$SR experiment}:  In a ${\mu}$SR experiment nearly 100 ${\%}$ spin-polarized muons ${\mu}$$^{+}$ are implanted into the sample one at a time. The positively charged ${\mu}$$^{+}$ thermalize at interstitial lattice sites, where they act as magnetic microprobes. In a magnetic material the  muon spin precesses in the local field $B_{\rm \mu}$ at the muon site with the Larmor frequency ${\nu}_{\rm \mu}$ = $\gamma_{\rm \mu}$/(2${\pi})$$B_{\rm \mu}$ (muon gyromagnetic ratio $\gamma_{\rm \mu}$/(2${\pi}$) = 135.5 MHz T$^{-1}$). Using the $\mu$SR technique, important length scales of superconductors can be measured and calculated, namely the magnetic penetration depth $\lambda$ and the coherence length $\xi$. If a Type II superconductor is cooled below $T_{\rm c}$ in an applied magnetic field ranging between the lower ($H_{c1}$) and the upper ($H_{c2}$) critical fields, a vortex lattice is formed which in general is incommensurate with the crystal lattice, with vortex cores separated by much larger distances than those of the crystallographic unit cell. Because the implanted muons stop at given crystallographic sites, they will randomly probe the field distribution of the vortex lattice. Such measurements need to be performed in a field applied perpendicular to the initial muon spin polarization (so-called TF configuration). \\

\textbf{Transverse Field $\mu$SR}:
We performed transverse field $\mu$SR (TF-$\mu$SR) experiments in the superconducting state to probe the superfluid density (see Eq. 11). Fits to the TF-$\mu$SR data in the normal state were performed using the following equation:
\begin{equation}
	A_{\rm TF}=A_0 \left[ \cos\left(2\pi\nu t + \frac{\pi \varphi}{180}\right) \times e^{-\frac{1}{2} (\sigma t)^2} \right]
\end{equation}
which describes the field distribution felt by the muons which stop in the sample \cite{Suter}. Here, the $A_0$ is the maximal value of the asymmetry; the Larmor frequency at which the muon spins are oscillating is described by $\nu$; the phase shift is captured by $\varphi$ and should be near 0$^\circ$; and each component can be represented by a Gaussian distribution of width $\sigma/\gamma_{\mu}$. In the normal state, the field distribution is fairly narrow and centered around the applied magnetic field (in this case, 70~mT, 20~mT, or 10~mT depending on the sample $H_{\rm c2}$; see Fig.~S5). In the superconducting state, however, the muon ensemble stopped in the superconducting volume fraction experiences a diamagnetic shift relative to the applied field as well as line broadening due to the increased distribution of local magnetic fields felt by the implanted muons. As the TF-$\mu$SR measurements were performed on powder samples, the field distribution in the superconducting component is symmetric due to powder averaging. However, as the superconducting volume fraction decreases with increasing dopant concentration, a portion of the sample remains in the normal state; this requires a multi-component Gaussian fit of the data (see Fig. S~6(b), (d), (f), (h), and (k)), using the following equation:
\begin{equation}
	A_{\rm TF, SC}=
	%A_1\left[ \cos\left(2\pi\nu_1 t + \frac{\pi \varphi_1}{180}\right) \times e^{-\frac{1}{2} (\sigma_1 t)^2} \right] + A_2 \left[ \cos\left(2\pi\nu_2 t + \frac{\pi \varphi_2}{180}\right) \times e^{-\frac{1}{2} (\sigma_2 t)^2}  \right]
	\sum_{j=1... n}A_j\left[ \cos\left(2\pi\nu_j t + \frac{\pi \varphi_j}{180}\right) \times e^{-\frac{1}{2} (\sigma_j t)^2} \right]
\end{equation}
where $j=1$ denotes the contribution to the first Gaussian distribution and $j=n$ denotes the $n$th Gaussian component \cite{Suter}. We found two components sufficient to describe all TF-$\mu$SR data taken in the superconducting state. When performing the fits to the data in the time domain, the asymmetry of the two components was fixed at the base temperature. The normal volume fraction was determined by the weight of the narrower Gaussian component centered at higher field values, while the superconducting volume fraction was taken from the broader Gaussian component experiencing a diamagnetic shift. The Larmor frequency and relaxation rate of each Gaussian component was then allowed to vary with respect to temperature. Examples of single- (normal state, in red) and two-component (SC state, in blue) Gaussian fits to the data may be seen in the Supplementary Information in Fig. S6, in both the time (Fig.~S6(a), (c), (e), (g) and (i)) and frequency (Fig.~S6(b), (d), (f), (h) and (k)) domains. 

The temperature dependence of the superconducting muon spin depolarization rate, $\sigma_{\rm SC}$, is shown in Fig.~3a. In order to investigate the symmetry of the SC gap, we note that ${\lambda}(T)$ is related to the depolarization rate ${\sigma}_{{\rm SC}}(T)$ in the presence of a perfect triangular vortex lattice with $H_{\rm app} \ll H_{\rm c2}$ by the equation \cite{Brandt}: 
\begin{equation}
	\frac{\sigma_{sc}(T)}{\gamma_{\mu}}=0.06091\frac{\Phi_{0}}{\lambda^{2}(T)},
\end{equation}
where ${\gamma_{\mu}}$ is the gyromagnetic ratio of the muon and ${\Phi}_{{\rm 0}}$ is the magnetic-flux quantum. The temperature dependence of the superfluid density $\lambda^{-2}(T)$ was then fitted with nodeless ($s$-wave) and nodal ($d$-wave) models to determine the superconducting gap symmetry.\\

%%%%%%%%%%%%%%%%%%%%%%%%%%%%%%%%%%%%%%%%%%%%%%%%%%%%%%%%%%%%%%
\begin{figure}[t!]
	\centering
	\includegraphics[width=0.90\linewidth]{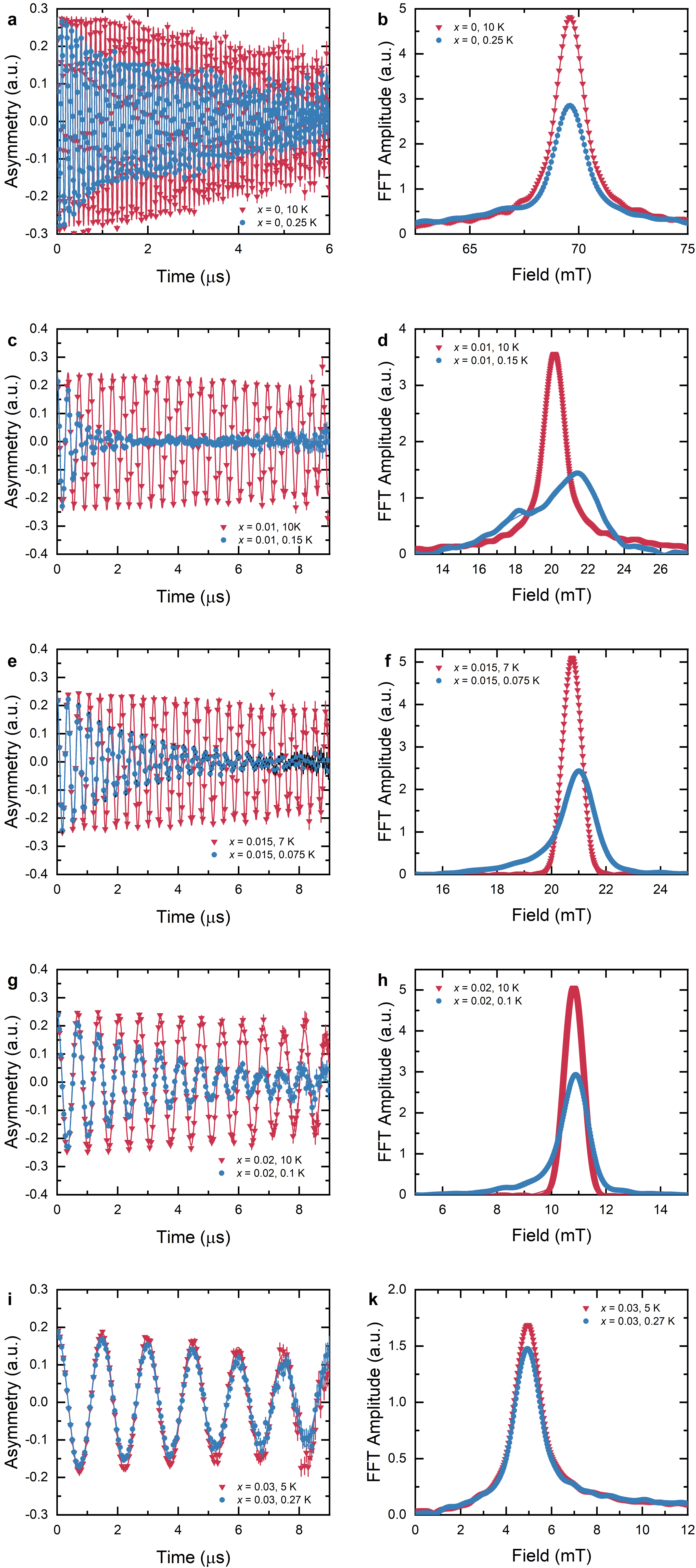}
	%\vspace{-2.8cm}
	\caption{{\bf $\mu$SR in the superconducting state of La(Ru$_{1-x}$Fe$_x$)$_3$Si$_2$.} $\mu$SR time spectra recorded in transverse field (TF) conditions in panels (a), (c), (e), (g) and (i), and their Fourier transforms are shown in panels (b), (d), (f), (h) and (k). The higher relaxation rate of the signal in the time spectra taken in the superconducting state can be seen as a characteristic line broadening in the Fourier transforms. The solid lines indicate fits to the data using Eq.s 9 and 10, and the error bars represent the standard deviation in $\simeq$5$\times$10$^6$ muon decay events.}
	\label{fig2}
\end{figure}
%%%%%%%%%%%%%%%%%%%%%%%%%%%%%%%%%%%%%%%%%%%%%%%%%%%%%%%%%%%%% 

\textbf{Longitudinal Field $\mu$SR}: Longitudinal field $\mu$SR (LF-$\mu$SR) experiments are performed with the magnetic field applied along the direction of the initial muon spin polarization. These measurements provide a powerful way to differentiate between static and dynamic/strongly fluctuating magnetism in materials where muon depolarization is detected by zero field (ZF)-$\mu$SR. 

%%%%%%%%%%%%%%%%%%%%%%%%%%%%%%%%%%%%%%%%%%%%%%%%%%%%%%%%%%%%%%
\begin{figure}[t!]
	\centering
	\includegraphics[width=1.0\linewidth]{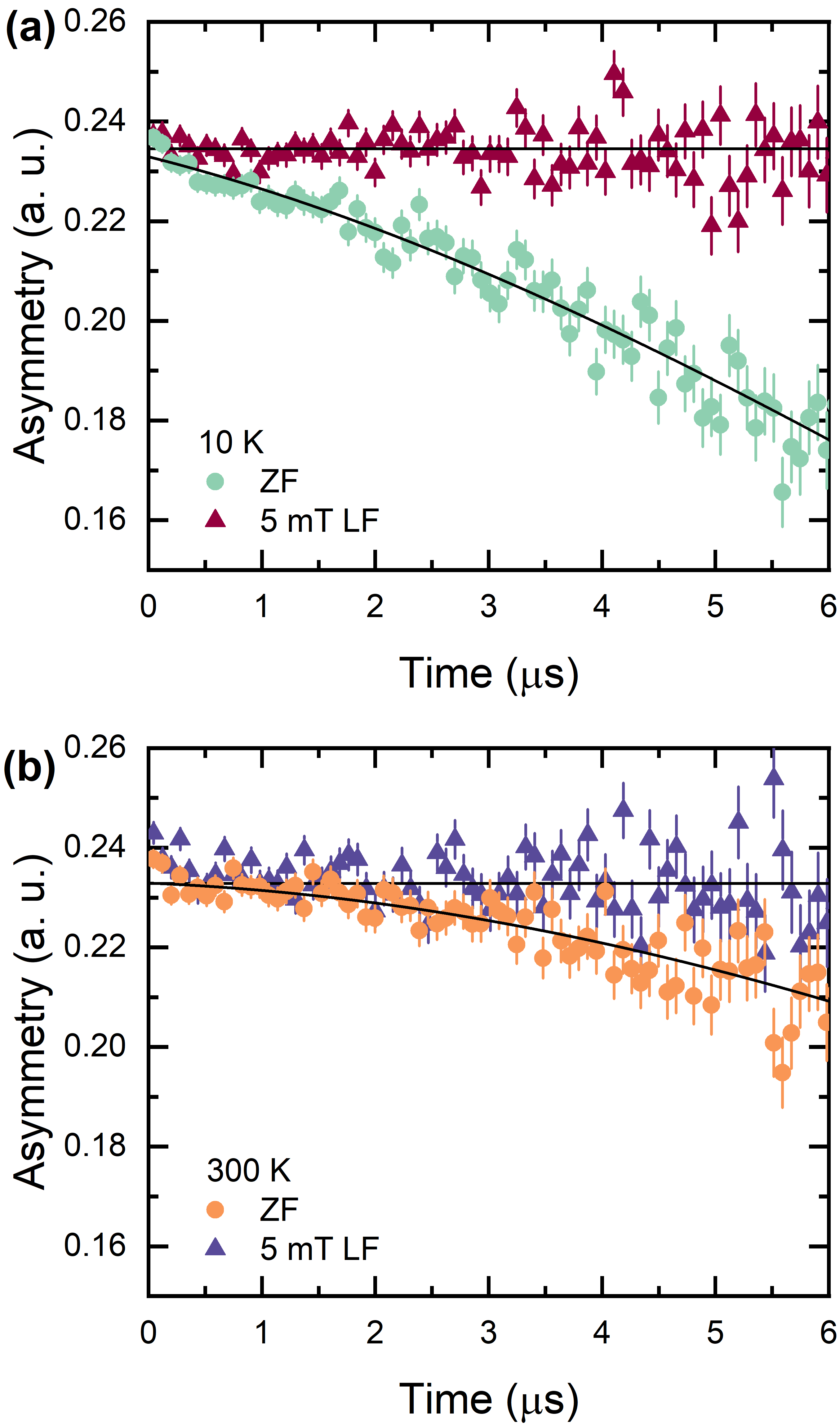}
	%\vspace{-2.8cm}
	\caption{{\bf Static nature of magnetism in LaRu$_3$Si$_2$.} $\mu$SR time spectra recorded in zero field (ZF) and a small longitudinal field of 5 mT, applied parallel to the muon spin polarization. The black lines indicate fit lines using Eq. 2 in the main text, and the error bars represent the standard deviation in $\simeq$6$\times$10$^6$ muon decay events. Panels (a) and (b) display ZF and LF-5~mT data collected at 10~K and 300~K, respectively.}
	\label{fig3}
\end{figure}
%%%%%%%%%%%%%%%%%%%%%%%%%%%%%%%%%%%%%%%%%%%%%%%%%%%%%%%%%%%%% 

%%%%%%%%%%%%%%%%%%%%%%%%%%%%%%%%%%%%%%%%%%%%%%%%%%%%%%%%%%%%%%
\begin{figure*}[t!]
	\centering
	\includegraphics[width=1.0\linewidth]{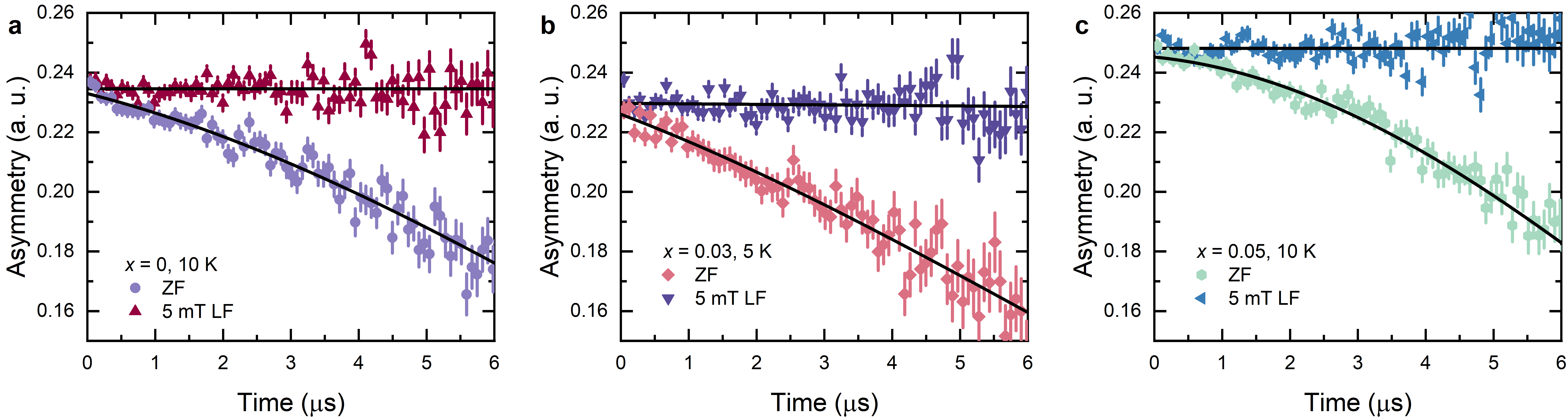}
	%\vspace{-2.8cm}
	\caption{{\bf Static nature of magnetism in LaRu$_3$Si$_2$.} $\mu$SR time spectra recorded in zero field (ZF) and a small longitudinal field of 5 mT, applied parallel to the muon spin polarization. The black lines indicate fit lines using Eq. 2 in the main text, and the error bars represent the standard deviation in $\simeq$6$\times$10$^6$ muon decay events. Panels (a), (b), and (c) display ZF and LF-5~mT data collected at low temperatures in the normal state for the undoped, $x$~=~0.03 and $x$~=~0.05 Fe-doped samples.}
	\label{fig4}
\end{figure*}
%%%%%%%%%%%%%%%%%%%%%%%%%%%%%%%%%%%%%%%%%%%%%%%%%%%%%%%%%%%%% 
On all samples analyzed by ZF-$\mu$SR, we also performed LF-$\mu$SR experiments to verify the static nature of the observed temperature-dependent muon depolarization rate. We were able to fully decouple (ie. recover the initial asymmetry) in weak fields $\simeq$5~mT at 10~K (see Fig. S\ref{fig3}, S\ref{fig4}), indicating the static nature of the weak magnetism observed in LaRu$_3$Si$_2$ and the Fe-doped variants. 

\clearpage

\textbf{Analysis of ${\lambda}(T)$}: 

\begin{table}
	\begin{tabular}{ c c c c c}
		\hline
		\multicolumn{5}{|c|}{Gap Symmetry Parameters} \\
		\hline
		$x_{\rm Fe}$ & Symmetry & $T_c$ [K] & $\sigma_{\rm SC}$ [$\mu$s$^{-1}$] & $\Delta$ [meV]\\
		\hline
		undoped &  $s$-wave & 6.299  & 1.825  & 1.144\\ %$\pm$0.010\\
		0.01 & $s$-wave & 4.699 & 2.057 &  0.604\\%$\pm$0.022\\
		0.015 & $d$-wave & 4.717 & 1.171 &  0.578\\ %$\pm$0.033\\
		0.02 & $d$-wave & 2.974 & 0.933 &  0.506\\ %$\pm$0.010\\
		0.03 & $d$-wave & 1.450 & 0.132 &  0.250\\ %$\pm$0.016\\
		\hline
	\end{tabular}
	\caption{ \textbf{Superconducting Gap Fits.}
		The parameters from the refinements to the superconducting gap symmetry using the musredit program \cite{Suter} are displayed above, indicating the gap symmetry function used (either $s$- or $d$-wave), the fitted critical temperature, the $\sigma_{\rm SC}$ (extrapolated to zero temperature) value and the superconducting gap value. }
\end{table} 

${\lambda}$($T$) was calculated within the local (London) approximation (${\lambda}$ ${\gg}$ ${\xi}$) by the following expression \cite{Suter,Tinkham}:
%%%%%%%%%%%%%%%%%%%%%%%%%%%%%%%%%%%%%%%%%%%%%%%%%%%%%%%%%%%%%%%%%%%%%
\begin{equation}
	\frac{\lambda^{-2}(T,\Delta_{0})}{\lambda^{-2}(0,\Delta_{0})}=
	1+\frac{1}{\pi}\int_{0}^{2\pi}\int_{\Delta(_{T,\phi})}^{\infty}(\frac{\partial f}{\partial E})\frac{E~dE~d\phi}{\sqrt{E^2-\Delta_i(T,\phi)^2}},
\end{equation}
%%%%%%%%%%%%%%%%%%%%%%%%%%%%%%%%%%%%%%%%%%%%%%%%%%%%%%%%%%%%%%%%%%
where $f=[1+\exp(E/k_{\rm B}T)]^{-1}$ is the Fermi function, ${\phi}$ is the angle along the Fermi surface, and ${\Delta}_{i}(T,{\phi})~=~{\Delta}_{0}~{\Gamma}~(T/T_{\rm c})~g({\phi}$)
(${\Delta}_{0}$ is the maximum gap value at $T=0$). The temperature dependence of the gap is approximated by the expression ${\Gamma}(T/T_{\rm c})~=~\tanh{\{}1.82~[1.018~(T_{\rm c}/T-1)]^{0.51}{\}}$,\cite{carrington} while $g({\phi}$) describes the angular dependence of the gap and it is replaced by 1 for an $s$-wave gap and ${\mid}\cos(2{\phi}){\mid}$ for a $d$-wave gap.

For the results of our gap symmetry fits, see Table II. \\


\newpage 

\begin{thebibliography}{150}


\bibitem{GuguchiaCSS} Z. Guguchia, J. Verezhak, D. Gawryluk, S. Tsirkin, J.-X. Yin, I. Belopolski, H. Zhou, G. Simutis, S.-S. Zhang, T. Cochran et al.
Tunable anomalous Hall conductivity through volume-wise magnetic competition in a topological kagome magnet.
Nature Communications \textbf{11}, 559 (2020).

\bibitem{JXYin2} J-X. Yin, S. Zhang, H. Li, K. Jiang, G. Chang, B. Zhang, B. Lian, C. Xiang, I. Belopolski, H. Zheng et al.
Giant and anisotropic spin-orbit tunability in a strongly correlated kagome magnet. 
Nature \textbf{562}, 91 (2018).

\bibitem{MielkeKVS} C. Mielke III, D. Das, J.-X. Yin, H. Liu, R. Gupta, Y.-X. Jiang, M. Medarde, X. Wu, H. Lei, J. Chang et al.
Time-reversal symmetry-breaking charge order in a kagome superconductor.
Nature \textbf{602}, 245 (2022).

\bibitem{JiangpingHu} K. Jiang, T. Wu, J.-X. Yin, Z. Wang, M.Z. Hasan, S.D. Wilson, X. Chen, \& J. Hu.
Kagome superconductors $A$V$_{3}$Sb$_{5}$ ($A$=K, Rb, Cs).
Nat. Sci. Rev. \textbf{10}, nwac199 (2023).

\bibitem{TNeupert} T. Neupert, M.M. Denner, J.-X. Yin, R. Thomale, \& M.Z. Hasan.
Charge order and superconductivity in kagome materials.
Nature Physics {\bf 18}, 137-143 (2022).

\bibitem{MielkeTMS} C. Mielke III, W. Ma, V. Pomjakushin, O. Zaharko, S. Sturniolo, X. Liu, V. Ukleev, J. White, J.-X. Yin, S. Tsirkin et al.
Low-temperature magnetic crossover in the topological kagome magnet TbMn$_6$Sn$_6$.
Communications Physics \textbf{5}, 107 (2022).

\bibitem{BOrtizCVS} B. Ortiz, S. Teicher, Y. Hu, J. Zuo, P. Sarte, E. Schueller, A. Abeykoon, M.  Krogstad, S. Rosenkranz, R. Osborn et al.
CsV$_3$Sb$_5$: A Z$_2$ Topological Kagome Metal with a Superconducting Ground State.
Phys. Rev. Lett. \textbf{125}, 247002 (2020).

\bibitem{YJiang} Y.-X. Jiang, J.-X. Yin, M. Denner, N. Shumiya, B. Ortiz, G. Xu, Z. Guguchia, J. He, M. Hossain, X. Liu et al.
Unconventional chiral charge order in kagome superconductor KV$_3$Sb$_5$.
Nature Materials \textbf{20}, 1353-1357 (2021).

\bibitem{KagomeReview} J.-X. Yin, B. Lian, \& M.Z. Hasan.
Topological kagome magnets and superconductors.
Nature \textbf{612}, 647-657 (2022).

\bibitem{Mazin} N. Ghimire \& I. Mazin.
Topology and correlations on the kagome lattice.
Nature Materials \textbf{19}, 137-138(2020).

\bibitem{LYe} L. Ye, M. Kang, J. Liu, F. von Cube, C. Wicker, T. Suzuki, C. Jozwiak, A. Bostwick, E. Rotenberg, D. Bell et al. 
Massive Dirac fermions in a ferromagnetic kagome metal. 
Nature \textbf{555}, 638-642 (2018).

\bibitem{Kiesel2013} M. Kiesel, C. Platt, \& R. Thomale. 
Unconventional Fermi Surface Instabilities in the Kagome Hubbard Model.
Phys. Rev. Lett. \textbf{110}, 126405 (2013). 

\bibitem{Nandkishore} Y.-P. Lin \& R. Nandkishore.
Complex charge density waves at Van Hove singularity on hexagonal lattices: Haldane-model phase diagram and potential realization in the kagome metals $A$V$_3$Sb$_5$ ($A=$ K, Rb, Cs).
Phys. Rev. B \textbf{104}, 045122 (2021).

\bibitem{Wang2013} W.-S. Wang, Z.-Z. Li, Y.-Y. Xiang, \& Q.-H. Wang
Competing electronic orders on kagome lattices at van Hove filling.
Phys. Rev. B {\bf 87}, 115135 (2013).

\bibitem{Qimiao} C. Setty, H. Hu, L. Chen, \& Q. Si.
Electron correlations and $T$-breaking density wave order in a $Z_{2}$ kagome metal,
arxiv:2105.15204 (2021).

\bibitem{MielkeCeRu2} C. Mielke III, H. Liu, D. Das, J.-X. Yin, L. Deng, J. Spring, R. Gupta, M. Medarde, C.-W. Chu, R. Khasanov et al. 
Local spectroscopic evidence for a nodeless magnetic kagome superconductor CeRu$_2$. 
J. Phys.: Condens. Matter \textbf{34}, 485601 (2022).

\bibitem{KhasanovCVS} R. Khasanov, D. Das, R. Gupta, C. Mielke III, M. Elender, Q. Yin, Z. Tu, C. Gong, H. Lei, E. Ritz et al.
Time reversal symmetry broken by charge order in CsV$_3$Sb$_5$.
Phys. Rev. Research \textbf{4}, 023244 (2022).

\bibitem{GuguchiaRVS} Z. Guguchia, C. Mielke III, D. Das, R. Gupta, J.-X. Yin, H. Liu, Q. Yin, M. Christensen, Z. Tu, C. Gong et al.
Tunable unconventional kagome superconductivity in charge ordered RbV$_3$Sb$_5$ and KV$_3$Sb$_5$.
Nature Communications \textbf{14}, 153 (2023).

\bibitem{GuptaCVS} R. Gupta, D. Das, C. Mielke III, E. Ritz, F. Hotz, Q. Yin, Z. Tu, C. Gong, H. Lei, T. Birol et al.
Two types of charge order with distinct interplay with superconductivity in the kagome material CsV$_3$Sb$_5$.
Communications Physics \textbf{5}, 232 (2022).

\bibitem{GuptaCVSnpj} R. Gupta, D. Das, C. Mielke III, Z. Guguchia, T. Shiroka, C. Baines, M. Bartkowiak, H. Luetkens, R. Khasanov, Q. Yin et al.
Microscopic evidence for anisotropic multigap superconductivity in the CsV$_3$Sb$_5$ kagome superconductor.
npj Quantum Materials \textbf{7}, 49 (2022).

\bibitem{MielkeLRS} C. Mielke III, Y. Qin, J.-X. Yin, H. Nakamura, D. Das, K. Guo, R. Khasanov, J. Chang, Z. Wang, S. Jia et al.
Nodeless kagome superconductivity in LaRu$_3$Si$_2$.
Phys. Rev. Materials \textbf{5}, 034803 (2021).

\bibitem{MielkeCDW} I. Plokhikh, C. Mielke III, H. Nakamura, V. Petricek, Y. Qin, V. Sazgari, J. Küspert, I. Bialo, S. Shin, O. Ivashko et al.
Charge order above room-temperature in a prototypical kagome superconductor La(Ru$_{1-x}$Fe$_x$)$_3$Si$_2$.
arxiv:2309.09255 (2023).

%%%%%
\bibitem{discoveryLRS} H. Barz. New ternary superconductors with silicon. Mater. Res. Bull. \textbf{15}, 1489 (1980); and J. M. Vandenberg \& H. Barz. \textit{ibid.} \textbf{15}, 1493 (1980).

\bibitem{Godart1987} C. Godart \& L. Gupta. 
Coexistence of superconductivity and spin glass freezing in La$_0.95$Gd$_0.05$Ru$_3$Si$_2$.
Phys. Lett. A \textbf{120}, 427 (1987).

\bibitem{Escorne1994} M. Escorne, A. Mauger, L. Gupta, \& C. Godart. 
Type-II superconductivity in a dilute magnetic system: La$_{1-x}Tm_x$Ru$_3$Si$_2$.
Phys. Rev. B \textbf{49}, 12051 (1994).
%%%%%

\bibitem{SLi2012} S. Li, J. Tao, X. Wan, X. Ding, H. Yang, \& H.-H. Wen.
Distinct behaviors of suppression to superconductivity in LaRu$_3$Si$_2$ induced by Fe and Co dopants.
Phys. Rev. B \textbf{86}, 024513 (2012).

\bibitem{BLi2016} B. Li, S. Li, \& H.-H. Wen.
Chemical doping effect in the LaRu$_3$Si$_2$ superconductor with a kagome lattice. 
Phys. Rev. B \textbf{94}, 094523 (2016).

\bibitem{SChakrabortty2023} S. Chakrabortty, R. Kumar, \& N. Mohapatra.
Effect of tunable spin-orbit coupling on the superconducting properties of LaRu$_3$Si$_2$ containing kagome-honeycomb layers.
Phys. Rev. B \textbf{107}, 024503 (2023). 

\bibitem{SLi2011} S. Li, B. Zeng, X. Wan, J. Tao, F. Han, H. Yang, Z. Wang, \& H.-H. Wen.
Anomalous properties in the normal and superconducting states of LaRu$_3$Si$_2$.
Phys. Rev. B \textbf{84}, 214527 (2011).

\bibitem{Guguchia2016} Z. Guguchia, R. Khasanov, Z. Bukowski, F. von Rohr, M. Medarde, P. K. Biswas, H. Luetkens, A. Amato, \& E. Morenzoni. 
Probing the pairing symmetry in the over-doped Fe-based superconductor Ba$_{0.35}$Rb$_{0.65}$Fe$_2$As$_2$ as a function of hydrostatic pressure. 
Phys. Rev. B \textbf{93}, 094513 (2016). 

\bibitem{Budko2009}  S. Bud'ko, N. Ni, \& P. Canfield.
Jump in specific heat at the superconducting transition temperature in Ba(Fe$_{1-x}$Co$_x$)$_2$As$_2$ and Ba(Fe$_{1-x}$Ni$_x$)$_2$As$_2$ single crystals. 
Phys. Rev. B \textbf{79}, 220516(R) (2009). 

\bibitem{Suter} A. Suter \& B.M. Wojek. 
Musrfit: a free platform-independent framework for $\mu$SR data analysis.
Physics Procedia \textbf{30}, 69 (2012).

\bibitem{Toyabe} R. Kubo \& T. Toyabe. Magnetic Resonance and Relaxation (North Holland, Amsterdam, 1967).

\bibitem{SuppMat} For Supplementary Information, see below for additional details related to synthesis, magnetization, heat capacity, and $\mu$SR experiments, as well as fitting procedures and parameters. The Supplementary Information contains references \cite{Brandt, Sonier, GuguchiaNature, SLi2012, BLi2016, SLi2011, Suter, Tinkham, carrington, Rodriguez, TShang2019, Uemura1, Uemura2, Shengelaya, GuguchiaMoTe2, GuguchiaNbSe2, YZhong} therein.

\bibitem{GuguchiaNature} Z. Guguchia, A. Amato, J. Kang, H. Luetkens, P. K. Biswas, G. Prando, F. von Rohr, Z. Bukowski, A. Shengelaya, H. Keller et al.
Direct evidence for the emergence of a pressure induced nodal superconducting gap in the iron-based superconductor Ba$_{0.65}$Rb$_{0.35}$Fe$_{2}$As$_{2}$.
Nature Communications \textbf{6}, 8863 (2015).

\bibitem{Fernandes2012} S. Maiti, R. Fernandes, \& A. Chubukov, 
Gap nodes induced by coexistence with antiferromagnetism in iron-based superconductors. 
Phys. Rev. B \textbf{85}, 144527 (2012).


\bibitem{Brandt} E.H. Brandt.  
Flux distribution and penetration depth measured by muon spin rotation in high-$T_{{\rm c}}$ superconductors. $Phys.~Rev.~B$ \textbf{37}, 2349 (1988).

\bibitem{Sonier} J. E. Sonier, J.H. Brewer, \& R.F. Kiefl. 
${\mu}$SR studies of the vortex state in type-II superconductors. Rev. Mod. Phys. \textbf{72}, 769 (2000).

\bibitem{Tinkham} M. Tinkham, Introduction to Superconductivity,
$Krieger~Publishing~Company$, $Malabar,~Florida$, 1975.

\bibitem{carrington} A. Carrington, \& F. Manzano.
Magnetic penetration depth of MgB$_{2}$. 
$Physica~C$ \textbf{385}, 205 (2003).

\bibitem{Rodriguez} J. Rodriguez-Carvajal.
Recent Advances in Magnetic Structure Determination by Neutron Powder Diffraction.
Physica B \textbf{192}, 55 (1993).

\bibitem{TShang2019} T. Shang, A. Amon, D. Kasinathan, W. Xie, M. Bobnar, Y. Chen, A. Wang, M. Shi, M. Medarde, H. Yuan, \& T. Shiroka.
Enhanced $T_c$ and multiband superconductivity in the fully-gapped ReBe$_{22}$ superconductor.
New J. Phys. \textbf{21}, 073034 (2019).

\bibitem{Uemura1}  Y. Uemura, G. Luke, B. Sternlieb, J. Brewer, J. Carolan, W. Hardy, R. Kadono, J. Kempton, R. Kiefl, S. Kreitzman et al. 
Universal Correlations between $T_{\rm c}$ and $n_{s}$/$m^{*}$ (Carrier Density over Effective Mass) in High-$T_{\rm c}$ Cuprate Superconductors.
Phys. Rev. Lett. \textbf{62}, 2317 (1989).

\bibitem{Uemura2} Y. Uemura, A. Keren, L. Le, G. Luke, B. Sternlieb, W. Wu, J. Brewer, R. Whetten, S. Huang, S. Lin et al.
Magnetic-field penetration depth in K$_{3}$C$_{60}$ measured by muon spin relaxation.
Nature \textbf{352}, 605 (1991).

\bibitem{Shengelaya} A. Shengelaya, R. Khasanov, D. Eshchenko, D. Di Castro, I. Savi\'{c}, M. Park, K. Kim, S.-I. Lee, K. M\"{u}ller, \& H. Keller. 
Muon-Spin-Rotation Measurements of the Penetration Depth of the Infinite-Layer Electron-Doped Sr$_{0.9}$La$_{0.1}$CuO$_{2}$ Cuprate Superconductor.
Phys. Rev. Lett. \textbf{94}, 127001 (2005).

\bibitem{GuguchiaMoTe2} Z. Guguchia, F. von Rohr, Z. Shermadini, A. Lee, S. Banerjee, A. Wieteska, C. Marianetti, B. Frandsen, H. Luetkens, Z. Gong et al. 
Signatures of the topological $s^{+-}$ superconducting order parameter in the type-II Weyl semimetal $T_{d}$-MoTe$_{2}$.
Nature Communications \textbf{8}, 1082 (2017).

\bibitem{GuguchiaNbSe2} F. O. von Rohr, J.-C. Orain, R. Khasanov, C. Witteveen, Z. Shermadini, A. Nikitin, J. Chang, A. Wieteska, A. N. Pasupathy, M. Z. Hasan et al. Unconventional Scaling of the Superfluid Density with the Critical Temperature in Transition Metal Dichalcogenides.
Science Advances \textbf{5(11)}, eaav8465 (2019).

\bibitem{YZhong} Y. Zhong, J. Liu, X. Wu, Z. Guguchia, J.-X. Yin, A. Mine, Y. Li, S. Najafzadeh, D. Das, C. Mielke III et al. 
Nodeless electron pairing in CsV$_{3}$Sb$_{5}$-derived kagome superconductors.
Nature \textbf{617}, 488 (2023).


\end{thebibliography}

\clearpage

\begin{thebibliography}{150}
	
	
	%%%%%%%%%%%%%%%%SUPPLEMENTARY MATERIAL%%%%%%%%%%%%%%%%%
	\bibitem{Brandt} E.H. Brandt.  
	Flux distribution and penetration depth measured by muon spin rotation in high-$T_{{\rm c}}$ superconductors. $Phys.~Rev.~B$ \textbf{37}, 2349 (1988).
	
	\bibitem{Sonier} J. E. Sonier, J.H. Brewer, \& R.F. Kiefl. 
	${\mu}$SR studies of the vortex state in type-II superconductors. Rev. Mod. Phys. \textbf{72}, 769 (2000).
	
	\bibitem{GuguchiaNature} Z. Guguchia, A. Amato, J. Kang, H. Luetkens, P. K. Biswas, G. Prando, F. von Rohr, Z. Bukowski, A. Shengelaya, H. Keller et al. 
	Direct evidence for the emergence of a pressure induced nodal superconducting gap in the iron-based superconductor Ba$_{0.65}$Rb$_{0.35}$Fe$_{2}$As$_{2}$.
	Nature Communications \textbf{6}, 8863 (2015).
	
	\bibitem{SLi2012} S. Li, J. Tao, X. Wan, X. Ding, H. Yang, \& H.-H. Wen.
	Distinct behaviors of suppression to superconductivity in LaRu$_3$Si$_2$ induced by Fe and Co dopants.
	Phys. Rev. B \textbf{86}, 024513 (2012).
	
	\bibitem{BLi2016} B. Li, S. Li, \& H.-H. Wen.
	Chemical doping effect in the LaRu$_3$Si$_2$ superconductor with a kagome lattice. 
	Phys. Rev. B \textbf{94}, 094523 (2016).
	
	\bibitem{SLi2011} S. Li, B. Zeng, X. Wan, J. Tao, F. Han, H. Yang, Z. Wang, \& H.-H. Wen.
	Anomalous properties in the normal and superconducting states of LaRu$_3$Si$_2$.
	Phys. Rev. B \textbf{84}, 214527 (2011).
	
	\bibitem{Suter} A. Suter \& B.M. Wojek. 
	Musrfit: a free platform-independent framework for $\mu$SR data analysis.
	Physics Procedia \textbf{30}, 69 (2012).
	
	\bibitem{Tinkham} M. Tinkham, Introduction to Superconductivity,
	$Krieger~Publishing~Company$, $Malabar,~Florida$, 1975.
	
	\bibitem{carrington} A. Carrington, \& F. Manzano.
	Magnetic penetration depth of MgB$_{2}$. 
	$Physica~C$ \textbf{385}, 205 (2003).
	
	\bibitem{Rodriguez} J. Rodriguez-Carvajal.
	Recent Advances in Magnetic Structure Determination by Neutron Powder Diffraction.
	Physica B \textbf{192}, 55 (1993).
	
	\bibitem{TShang2019} T. Shang, A. Amon, D. Kasinathan, W. Xie, M. Bobnar, Y. Chen, A. Wang, M. Shi, M. Medarde, H. Yuan, \& T. Shiroka.
	Enhanced $T_c$ and multiband superconductivity in the fully-gapped ReBe$_{22}$ superconductor.
	New J. Phys. \textbf{21}, 073034 (2019).
	
	\bibitem{Uemura1} Y. Uemura, G. Luke, B. Sternlieb, J. Brewer, J. Carolan, W. Hardy, R. Kadono, J. Kempton, R. Kiefl, S. Kreitzman et al. Universal Correlations between $T_c$ and n$_s$/m$_*$ (Carrier Density over Effective Mass) in High-$T_c$ Cuprate Superconductors. Phys. Rev. Lett. \textbf{62}, 2317 (1989).
	
	\bibitem{Uemura2} Y. Uemura, A. Keren, L. Le, G. Luke, B. Sternlieb, W. Wu, J. Brewer, R. Whetten, S. Huang, S. Lin et al. Magnetic-field penetration depth in K$_3$C$_60$ measured by muon spin relaxation. Nature \textbf{352}, 605 (1991).
	
	\bibitem{Shengelaya} A. Shengelaya, R. Khasanov, D. Eshchenko, D. Di Castro, I. Savi\'{c}, M. Park, K. Kim, S.-I. Lee, K. M\"{u}ller, \& H. Keller. Muon-Spin-Rotation Measurements of the Penetration Depth of the Infinite-Layer Electron-Doped Sr$_{0.9}$La$_{0.1}$CuO$_2$ Cuprate Superconductor. Phys. Rev. Lett. \textbf{94}, 127001 (2005).
	
	\bibitem{GuguchiaMoTe2} Z. Guguchia, F. von Rohr, Z. Shermadini, A. Lee, S. Banerjee, A. Wieteska, C. Marianetti, B. Frandsen, H. Luetkens, Z. Gong et al. 
	Signatures of the topological $s^{+-}$ superconducting order parameter in the type-II Weyl semimetal $T_{d}$-MoTe$_{2}$.
	Nature Communications \textbf{8}, 1082 (2017).
	
	\bibitem{GuguchiaNbSe2} F. O. von Rohr, J.-C. Orain, R. Khasanov, C. Witteveen, Z. Shermadini, A. Nikitin, J. Chang, A. Wieteska, A. N. Pasupathy, M. Z. Hasan et al. Unconventional Scaling of the Superfluid Density with the Critical Temperature in Transition Metal Dichalcogenides.
	Science Advances \textbf{5(11)}, eaav8465 (2019).
	
	\bibitem{YZhong} Y. Zhong, J. Liu, X. Wu, Z. Guguchia, J.-X. Yin, A. Mine, Y. Li, S. Najafzadeh, D. Das, C. Mielke III et al. Nodeless electron pairing in CsV$_{3}$Sb$_{5}$-derived kagome superconductors.
	Nature \textbf{617}, 488 (2023).
	
	
\end{thebibliography}
\end{document}